\documentclass[a4paper,10pt,twoside]{article}
\usepackage[latin1]{inputenc}
\usepackage{rotating}
\usepackage{graphicx}
\usepackage{epsfig}
\usepackage{subfigure}
\usepackage{fancyhdr}
\usepackage{placeins}
\usepackage[round]{natbib}

\newcommand{\pdfauthor}{F. Fabián Rosales Ortega}
\newcommand{\pdftitle}{PINGSoft documentation}
\newcommand{\pdfsubject}{PINGSoft documentation}

\usepackage[pdftex,pdfpagemode=UseOutlines,breaklinks=true]{hyperref}

\hypersetup{citecolor={blue},linkcolor={blue},urlcolor={blue}}     

\hypersetup{
  colorlinks=true,             
  bookmarksnumbered=true,      
  bookmarksopen=false,         
  pdftitle={\pdftitle},        
  pdfauthor={\pdfauthor},      
  pdfsubject={\pdfsubject},    
  pdfpagelayout=TwoPageRight,  
  pdfdisplaydoctitle           
}

\usepackage[left=2.5cm,right=2.5cm,top=2.3cm,bottom=2.3cm,head=1cm,foot=1cm]{geometry}

\begin{document}

\pdfbookmark[1]{Front matter}{title}
\thispagestyle{empty}
\begin{center}

  \sf

  \vspace*{0.05\textheight}

  \includegraphics[width=7cm]{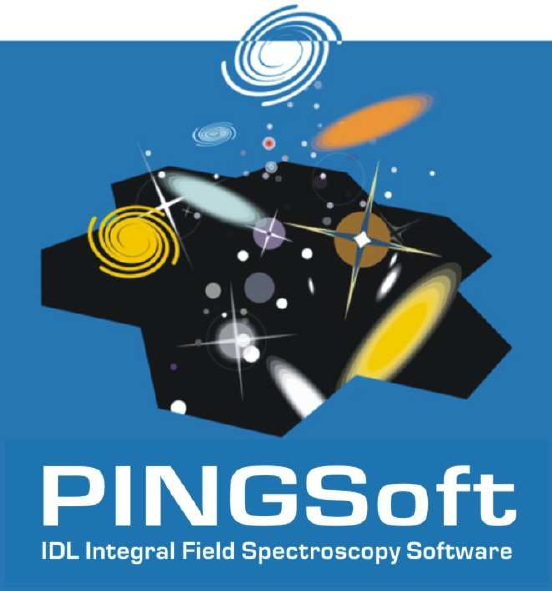}

  \vspace*{0.05\textheight}


  {\Huge PINGSoft 2:}\\
  \vspace{0.8\baselineskip}
  {\Huge an IDL Integral Field Spectroscopy Software}\\

  \vspace*{0.2\textheight}

  \vspace{0.5\baselineskip}
  {\Large F. Fabián Rosales-Ortega}\\
  \vspace{0.8\baselineskip}

  {\small Departamento de Física Teórica, Universidad Autónoma de Madrid, Spain}\\
  {\small Instituto Nacional de Astrofísica, Óptica y Electrónica, Mexico}\\

  \vspace{1cm}
  \href{mailto:frosales@inaoep.mx}{\large\tt frosales@inaoep.mx}\\
  \vspace{0.8\baselineskip}

  \vfill

  {\large November, 2012}
  
  \vspace{0.5\baselineskip}

\end{center}
\newpage






\pagestyle{plain}
\setcounter{page}{1}

\section*{What is {\sf PINGSoft}?}

The {\sf PINGS software}, or {\sf PINGSoft}, is a set of IDL routines designed
to visualise, manipulate and analyse in a simply way Integral Field
Spectroscopic (IFS) data regardless of the original instrument and spaxel
size/shape, it is able to run on practically any computer platform with
minimal library requirements \citep{RosalesOrtega:2011p3845}.\\

{\sf PINGSoft 2} is a relatively major upgrade with respect to the first
version. The overall functionality and layout have been improved, while the
command syntax has been simplified. This version includes new routines that
offer additional extraction options and some level of analysis. 
{\sf PINGSoft} includes basic tools to visualise spatially and spectrally 
IFS data, to extract regions of interest by hand or within a given geometric
aperture, to integrate the spectra within a given region, and to perform simple
analysis to the IFS data. Additionally, some miscellaneous codes useful for
generic tasks performed in astronomy and spectroscopy are also included.
{\sf PINGSoft} is optimised for a fast visualisation rendering, it can be
adapted to work with practically any IFS data/instrument, and is adapted to
work natively with the \href{http://califa.caha.es}{CALIFA survey} data.\\

\vspace{1cm}
\subsection*{New features}

\begin{itemize}

 \item A new Graphical User Interface (widget) for interactive visualisation
   of the spaxels and spectra of a 3D cube or RSS file.

  \item The data can be convolved with a full set of narrow and broad-band filters for
    visualization and/or analysis purposes. The filter used to visualize the data
    is shown on the spectral window.

  \item Elliptical apertures for spectra extraction are now supported (for any
    ellipticity, size and PA).

  \item Radial binning extraction with either fixed bins or based on a S/N floor.

  \item Spectra extraction and integration based on a user-given mask.

  \item Conic or hyperbolic aperture extractions for any PA, size and angle.

  \item Spectra integration based on S/N on continuum and/or emission line
    features.

  \item Voronoi binning based on the method developed by \citet{Cappellari:2003p4046}.

  \item Intrinsic velocity field correction using a wavelength cross-correlation.

  \item Furthermore, the {\sf PINGSoft} routines can now read 3D cubes or RSS files
    indistinctively. The syntax is much simpler, e.g. to load a RSS file the
    user only needs to include the name of the FITS file (and omit the position table) if the
    format is the following: {\tt OBJECT.fits}, {\tt OBJECT.pt.txt}, and both files reside
    on the same directory.

\end{itemize}

\vspace{2cm}

\noindent
{\large
In this document, we introduce the new visualisation tools of {\sf PINGSoft}.
We also list all the available routines for spectra extraction, analysis and
manipulation of IFS data. A detailed description of the remaining
routines can be found in the {\sf PINGSoft 2} User's Guide, together with
detailed installation instructions, all available at the project webpage:\\

\noindent
\url{http://califa.caha.es/pingsoft}
}
\newpage

\section*{The {\sf PINGSoft} integral field spectroscopy software}
\vspace{0.5cm}

All {\sf PINGSoft} routines are called via command lines in a terminal running
IDL. The syntax and online help for any program can be obtained by
entering the name of the procedure without any parameter or keyword, With
the exception of the visualisation widget: {\tt view\_ifs, /help}.\\

Additionally, in the {\sf PINGSoft} webpage you can download the 
\href{http://www.ast.cam.ac.uk/ioa/research/pings/media/pingsoft_examples.tar.gz}{\tt pingsoft\_examples/}
directory which includes some 3D cubes and RSS example files.
I recommend the user to download this example data and
follow the instructions in the {\tt README.pro} file in order to get a first
insight of the main {\sf PINGSoft} routines. All the example commands used in
this document can be found in the {\tt README.pro} file of the 
{\tt pingsoft\_examples/} directory.\\

\subsection*{\Large IFS visualisation}
\vspace{0.5cm}

\subsection*{\Large \texttt{view\_ifs}}
\addcontentsline{toc}{subsection}{\texttt{view\_ifs}}
\label{view_ifs}
\vspace{0.2cm}

This routine provides a spatial and spectral interactive visualisation widget
for 3D cubes and RSS IFS files. If the command is simply entered in the IDL
terminal:\\\vspace{-10pt}

{\small
\begin{verbatim}
IDL> view_ifs
\end{verbatim}
}

\noindent
it prompts for an input FITS file using a dialog window. Otherwise, the input
file can be passed directly to the command as the first
parameter:\\\vspace{-10pt}

{\small
\begin{verbatim}
IDL> view_ifs, 'OBJECT.fits'
\end{verbatim}
}

\noindent
The widget will be displayed automatically if the input file is a 3D FITS
cube. If the input FITS is a RSS file, the program will look in the same directory
for a position table named {\tt OBJECT.pt.txt} and will launch the widget if
the file exists. If this is not the case, the program will exit with an
error. The user can define the name of the corresponding position table using
the {\tt PT} parameter:\\\vspace{-10pt}

{\small
\begin{verbatim}
IDL> view_ifs, 'OBJECT.fits', PT='PosTable.txt'
\end{verbatim}
}
\vspace{0.5cm}

\begin{figure}[!t]
  \begin{center}
    \includegraphics[width=0.9\textwidth]{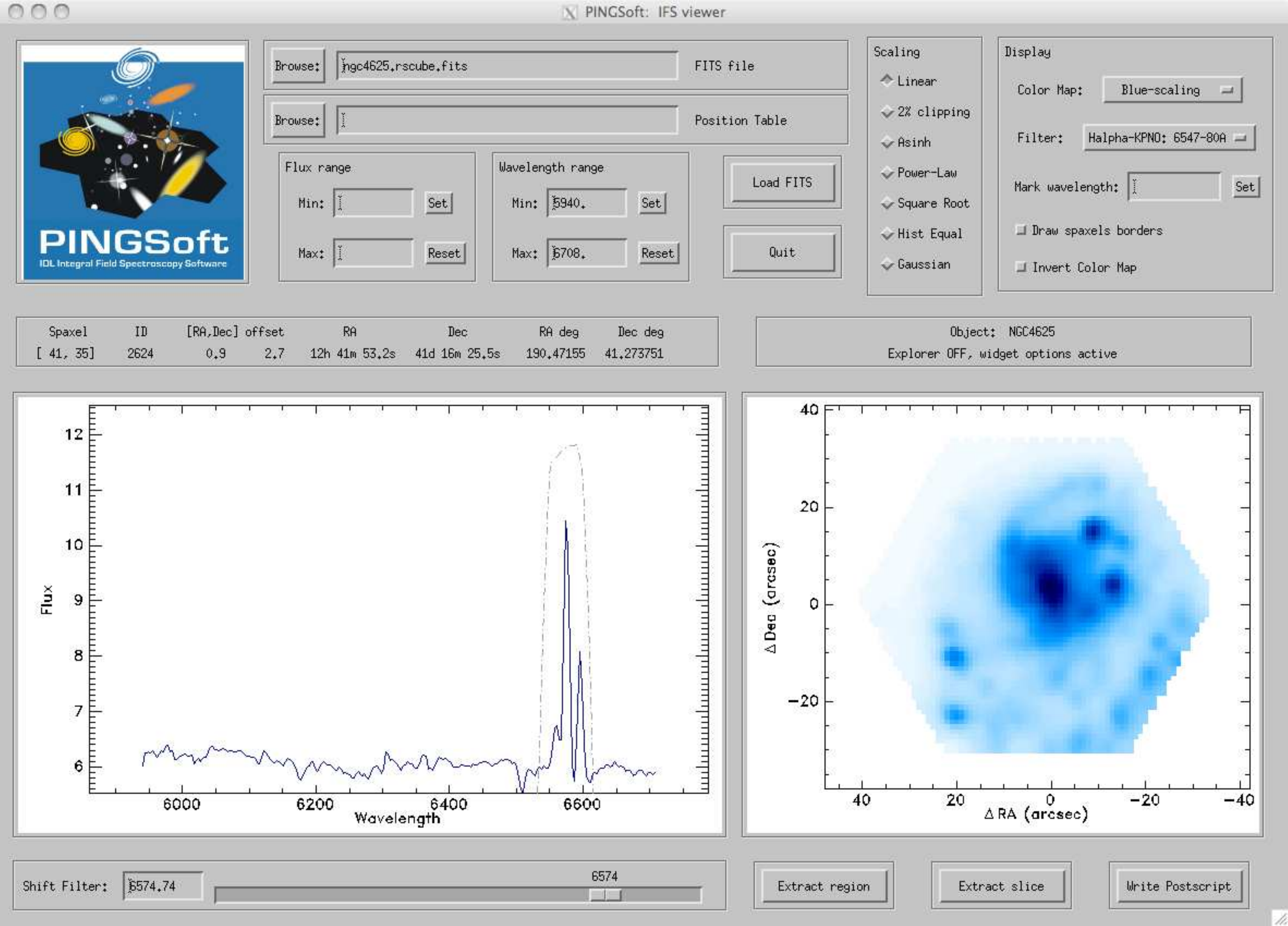}
    \caption[]
    {Screen shot of the visualisation widget launched by the {\tt view\_ifs}
      command, displaying the 3D cube of the galaxy NGC\,4625 included
      in the {\tt pingsoft\_examples/} directory.
      \label{fig:view_ifs}
    }
  \end{center}
\end{figure}

\noindent 
The {\tt view\_ifs} widget is shown in \autoref{fig:view_ifs}, displaying the
3D cube file {\tt ngc4625.rscube.fits} included in the 
{\tt pingsoft\_examples/} directory.
The widget displays two main panels, on the right a
visualisation of the spatial distribution of spaxels (or field-of-view,
FoV). The color-scaling corresponds to a narrow/broad-band image of a
transmission filter convolved with the data at a given wavelength
\footnote{~Default: H$\alpha$ narrow-band filter of 80~\AA\ FWHM with central
  wavelength at 6547 \AA, if this wavelength is outside the spectral range,
  the filter is shifted to the mean wavelength.} (shown as a dotted-curve in
the spectral window).
The spatial units are assumed to be arcseconds in a standard North (up) East
(left) configuration. The left panel shows the spectrum of the spaxel corresponding
to the position of the mouse, the wavelength range is extracted from the
information on the FITS header. The corresponding spaxel position,
ID\footnote{~In the IDL format, i.e. starting at zero} and offsets are shown
on the top of the spectral window. Optionally, if the WCS is included in the
FITS header, the RA and Dec are also shown in sexagesimal and degree units,
a mouse LEFT-click prints the same spaxel information on the IDL terminal
where the program was called.\\

\noindent
{\bf USAGE:} At first glance, the usage of the {\tt view\_ifs} widget may seem ``tricky'',
but it is easy to get used to: when the widget is launched, the user can
explore spatially and spectrally the IFS data but the widget options will be
{\em inactive}. To have access to the widget options, the user
needs to RIGHT-click with the mouse over the FoV panel. When the widget
options are changed the spectral explorer will become active again and the
widget options will be unavailable. To active/deactivate the widget options the
user only needs to RIGHT-click over the FoV to switch between the explorer
ON/OFF options. The {\em status} bar above the FoV panel (below the object
name) will indicate whether the explorer is active or not.\\

Several options are available to visualise the data, including different
intensity scalings, color maps, a set of different narrow and broad-band
filters in the optical to generate the visualisation in the FoV panel, the
choice to define the flux intensity and spectral ranges, to drawing the
contour of the spaxels, to invert the color-map, etc. Note that the central
wavelength of the filter used to display the data can be shifted to any
position along the spectral range, either by using the slider or by setting
the wavelength in the corresponding field. New FITS (position tables) files
can be loaded using the corresponding fields at the top-center of the widget,
and by pressing the ``Load FITS'' button.\\

\noindent
{\bf Extract region:}
By pressing this button, all the subsequent
LEFT-clicks over the FoV panel will mark and select the spaxels to be
extracted. When the program is terminated (by RIGHT-click) the following files
are created:

\vspace{0.2cm}
{\footnotesize
\begin{verbatim}
      Extracted RSS:   OBJECT_rss.fits    (Extracted RSS of the selected spaxels)
     Position table:   OBJECT_rss.pt.txt  (Position table of the new RSS file)
   Integrated ASCII:   OBJECT_integ.txt   (Integrated spectrum in ASCII format)
               FITS:   OBJECT_integ.fits  (Integrated spectrum in FITS format)
         Postscript:   OBJECT_integ.eps   (Postscript image of the integrated spectrum)
        IDL indices:   OBJECT_index.txt   (IDL indices of the selected spaxels)
\end{verbatim}
}
\vspace{0.2cm}

\noindent 
shown in the IDL terminal window, while the spectral panel will show the
integrated spectrum of the selected spaxels.\\

\noindent
{\bf Extract slice:}
Pressing this button will invoke the {\tt extract\_filter} command with the
filter and central wavelength parameters as the current values displayed in
the widget. This will create a FITS image file called {\tt OBJECT\_slice.fits}
as reported in the status bar, additional information will be shown in the IDL
terminal window. WARNING: This option is only available for 3D cubes and RSS with a
rectangular-contiguous grid.\\

\noindent
{\bf Write Postscript:}
Pressing this button will create an encapsulated Postscript image 
({\tt OBJECT\_FoV.eps}) of the FoV panel
with the current display options of the widget. The name of the file will be
reported in the status bar and terminal window.\\

\noindent
{\bf Mark wavelength:}
Use this field to enter one or several wavelengths at which a vertical line
will be drawn in the spectral panel. This option is useful when trying to
identify features at known wavelengths. The input formats can be of the
form:\\\vspace{-10pt}

{\footnotesize
\begin{verbatim}
Mark wavelength: 6563
                 5007, 6563
                 5895*1.002          (e.g. known redshifts)
                 [4310,5876]*1.0015
\end{verbatim}
}
\vspace{0.5cm}

\begin{figure}[!t]
  \begin{center}
    \includegraphics[width=0.9\textwidth]{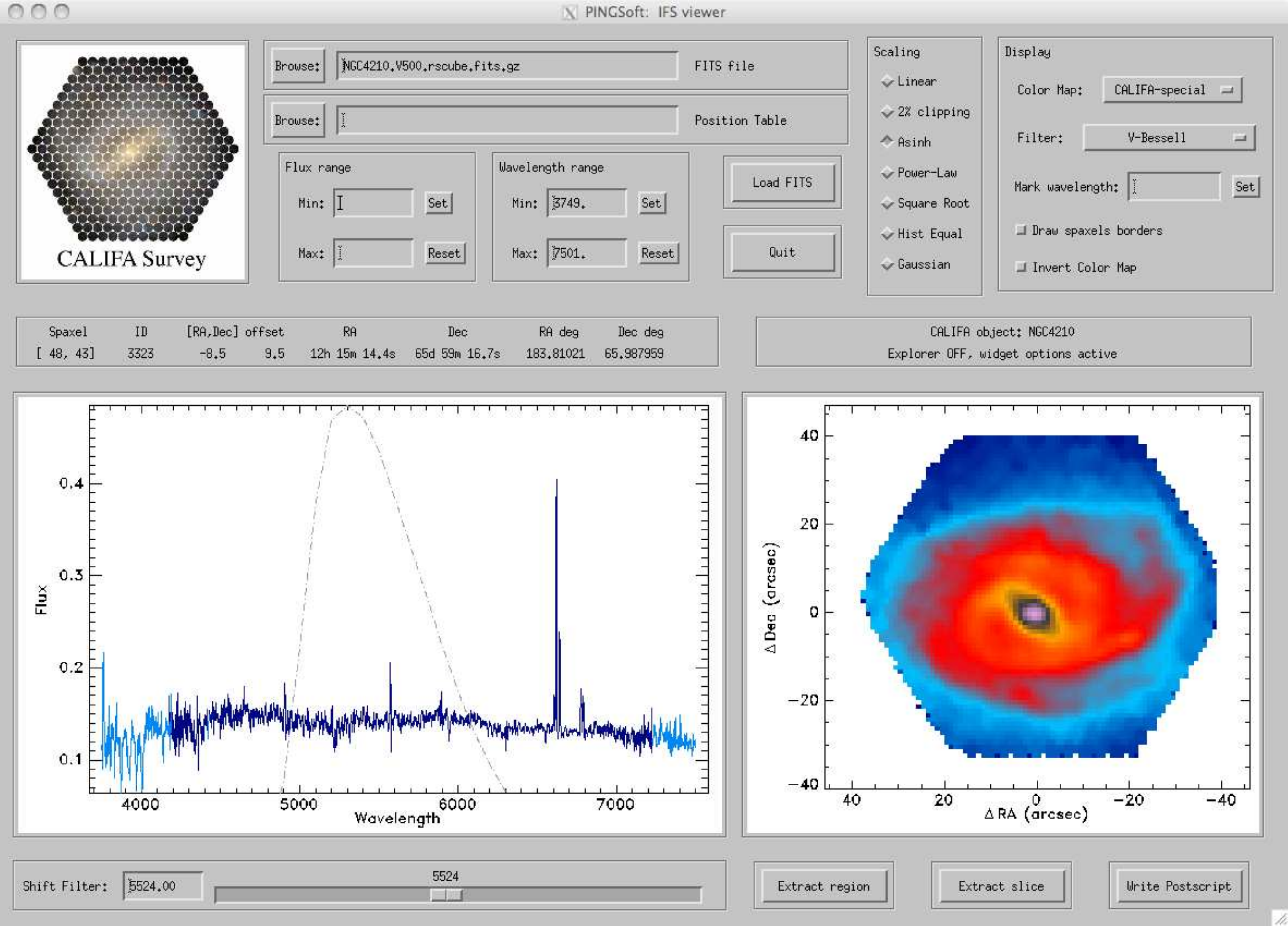}
    \caption[]
    {Screen shot of the visualisation widget for a CALIFA datacube, the
      bad-pixels are displayed in a light-blue color in the spectral panel.
      \label{fig:califa}
    }
  \end{center}
\end{figure}

\noindent
{\bf CALIFA data:}
When a CALIFA data file is loaded with {\tt view\_ifs}, the display color is
changed to the CALIFA-special color map, the routine identifies
automatically the different FITS extensions (HDUs) of the CALIFA format (for
both RSS and 3D cube versions).
The BADPIX extension is shown in the spectral panel simultaneously with
the flux data, the bad-pixels are displayed in a light-blue color as shown in
\autoref{fig:califa}.\\

\noindent
{\bf Size and resolution:}
The {\tt view\_ifs} widget may not display properly for screens with resolutions
lower than 1400$\times$800. In this case, the user can modify by hand the size
of the widget to fit their own screen resolution by editing the first entries
in the {\tt widget\_param.pro} routine.

\subsection*{\Large \texttt{view\_3D}}
\addcontentsline{toc}{subsection}{\texttt{view\_3D}}
\label{view_3d}
\vspace{0.2cm}

This routine is the command-line version of {\tt view\_ifs}, it provides 
a 2D interactive visualisation of the spaxels and spectra of a 3D cube or a
RSS file and its corresponding position table. However, the visualisation is
performed in two standard IDL windows (i.e. a lighter visualisation option). 
All display and interactive options are similar to {\tt view\_ifs} (with the
exception of the ``Extract slice'' option), the command is terminated by
RIGHT-click on the FoV window.\\

A mouse MIDDLE-button click is equivalent to the ``Extract region'' button in
the widget, it prompts in the IDL terminal for a PREFIX used to generate the new series
of files, all the subsequent LEFT-clicks over the FoV window
will mark and select the spaxels to be extracted. The extraction is performed
by RIGHT-click on the FoV window, the extracted files are displayed in the
terminal window, while the spectral window shows the integrated spectrum of
the selected spaxels.\\


{\bf Additional features:}

\begin{enumerate}

  \item The {\tt view\_3D} routine accepts the {\tt /LARGE} keyword, which
    displays a much larger FoV window.

  \item The user can specifying the FITS extension to read using the {\tt
      EXTENSION} keyword

  \item An optional output IDL structure can be obtained when spaxels are manually
    selected.

  \item The {\tt \_EXTRA} structure keyword can be set for user's defined
    specific graphics output, both for the IDL window or Postscript output, 
    e.g. {\tt \_EXTRA=\{title:'IFS cube',xrange:[-60,50]\}}

\end{enumerate}

\vspace{0.2cm}
\noindent {\sc Calling sequence:}

{\footnotesize
\begin{verbatim}
 view_3D, 'OBJECT.fits' [, OUT.str, PT='Ptable.txt', EXTENSION=extension, $
                        MIN_FLUX=min_flux, MAX_FLUX=max_flux, LMIN=lam_min, LMAX=lam_max, $
                        FILTER=filter, BAND=band, CT=ct, VLINE=vline, FONT=font, $
                        /CLIP, /GAMMA, /LOG, /ASINH, /SQRT, /HISTOGRAM, /GAUSSIAN, $
                        /PS, /DRAW, /LARGE, /NOBAND, _EXTRA={extra} ]
                             
 INPUTS:
       
         'OBJECT.fits':  String of the wavelength calibrated 3D cube or RSS FITS file.

 OPTIONAL KEYWORDS:
       
               OUT.str:  Output IDL structure (when spaxels are manually selected).

       PT='Ptable.txt':  Name of the position table in ASCII format for an input RSS file
                         in (North-East configuration).
                         NOTE: compulsory if not included in the default instruments/setups
                         or when the name is not in the 'OBJECT.pt.txt' format.

             EXTENSION:  Non-negative scalar integer specifying the FITS extension to read.
                         For example, specify EXTENSION = 1 to read the first FITS extension.   

          MIN/MAX_FLUX:  Minimum/maximum flux in the spectral window to be plot, if
                         not set these are floating values.

             LMIN/LMAX:  Defines the wavelength range on the spectral window, 
                         if not set values are taken from the FITS header.
                          
                FILTER:  Internal number of the narrow or broad-band filter used to
                         display the data. Available filters and corresponding
                         numbers can be obtained by typing:  IDL> pingsoft_filters
                         Default: 1 (Halpha KPNO-NOAO - CWL: 6547A  FWHM: 80)

                  BAND:  Central wavelength of the narrow or broad-band used to
                         display the data, i.e. shifts the band to the position
                         defined by the user (if within the spectral range).
                         Defaults: nominal central wavelength of the
                         corresponding filter (mean wavelength if outside the range).

                 VLINE:  Either a floating value or a vector of floating
                         numbers containing the lambda value at which a single or
                         several vertical lines will be drawn for reference
                         purposes, equivalent to "Mark wavelength" in the VIEW_IFS widget,
                         e.g. VLINE=6500 or VLINE=[4200,5400,6700].

                    CT:  IDL Color Table used to display the data, (default ct=1, BLUE/WHITE).
                     
                   /PS:  Writes an encapsulated Postscript file of the spaxels visualisation.

                  FONT:  Postscript IDL font to be used when /PS is set. Default: 12 (Helvetica)
    
                 /DRAW:  Draws the contours of the spaxels.

                /LARGE:  Displays a larger window for the spatial distribution
                         of spaxels (right window). 

               /NOBAND:  The narrow/broad band is not drawn in the spectral window.

                _EXTRA:  Structure with the _EXTRA tags for user's defined graphics output,
                         e.g. _EXTRA={title:'IFS cube'}

  Intensity Scalings:

               Default:  LINEAR, displays the range of intensities using a linear min/max scaling.

                 /CLIP:  A histogram stretch, with a 2% of pixels clipped at both the top and bottom.

                /GAMMA:  Displays the range of intensities using a Power-law (gamma) scaling.

                  /LOG:  Displays the range of intensities using a logarithmic scaling.

                /ASINH:  Displays the range of intensities using an inverse hyperbolic sine function scaling.

                 /SQRT:  Displays a linear stretch of the square root histogram of the image values.

            /HISTOGRAM:  Displays a linear stretch of the histogram equalized image histogram.

             /GAUSSIAN:  The scaling is performed by applying a Gaussian normal function to the image histogram.
\end{verbatim}
}
\vspace{0.3cm}

\begin{figure}[!t]
  \begin{center}
    \includegraphics[width=0.47\textwidth]{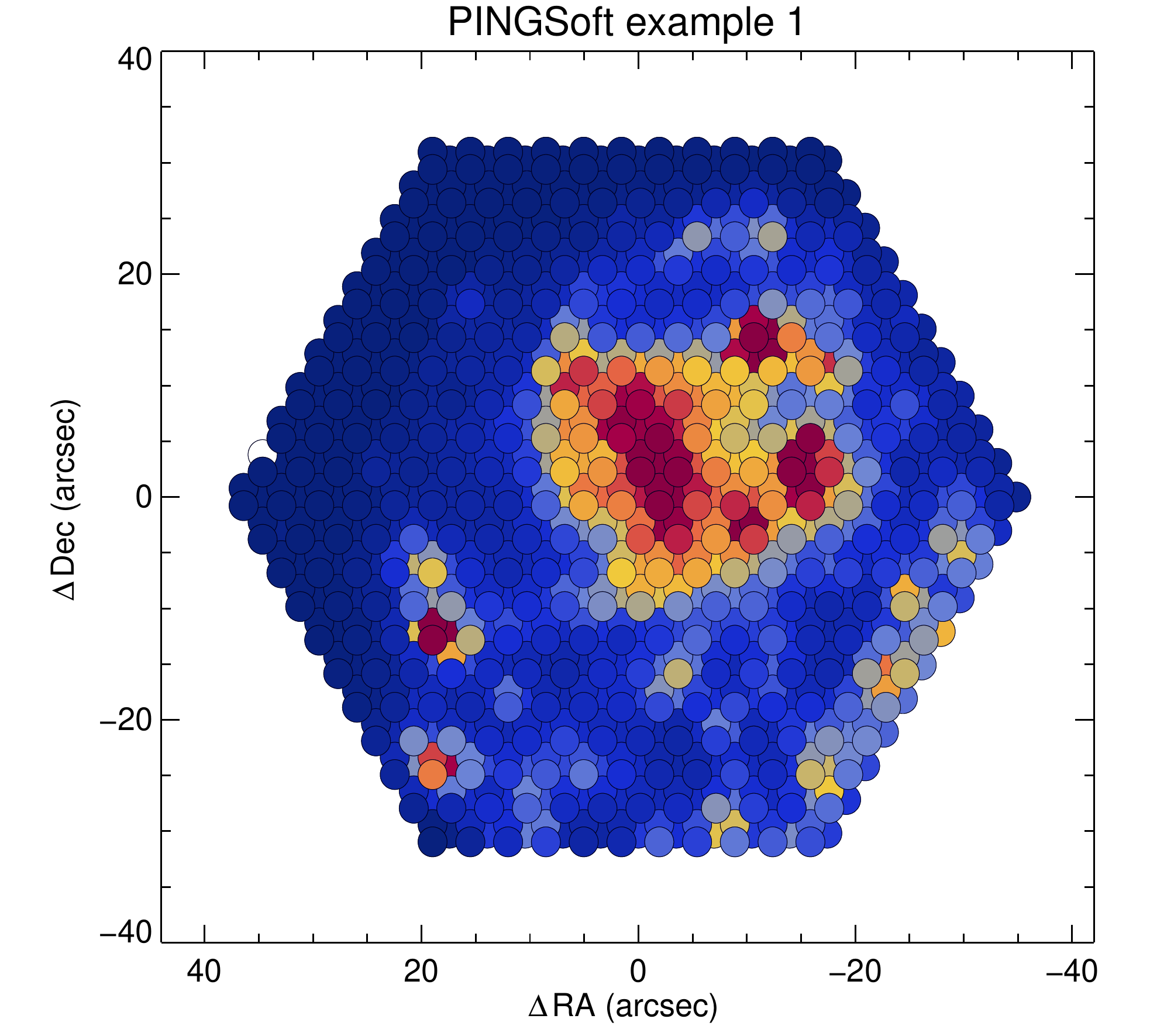}\hspace{0.2cm}
    \includegraphics[width=0.47\textwidth]{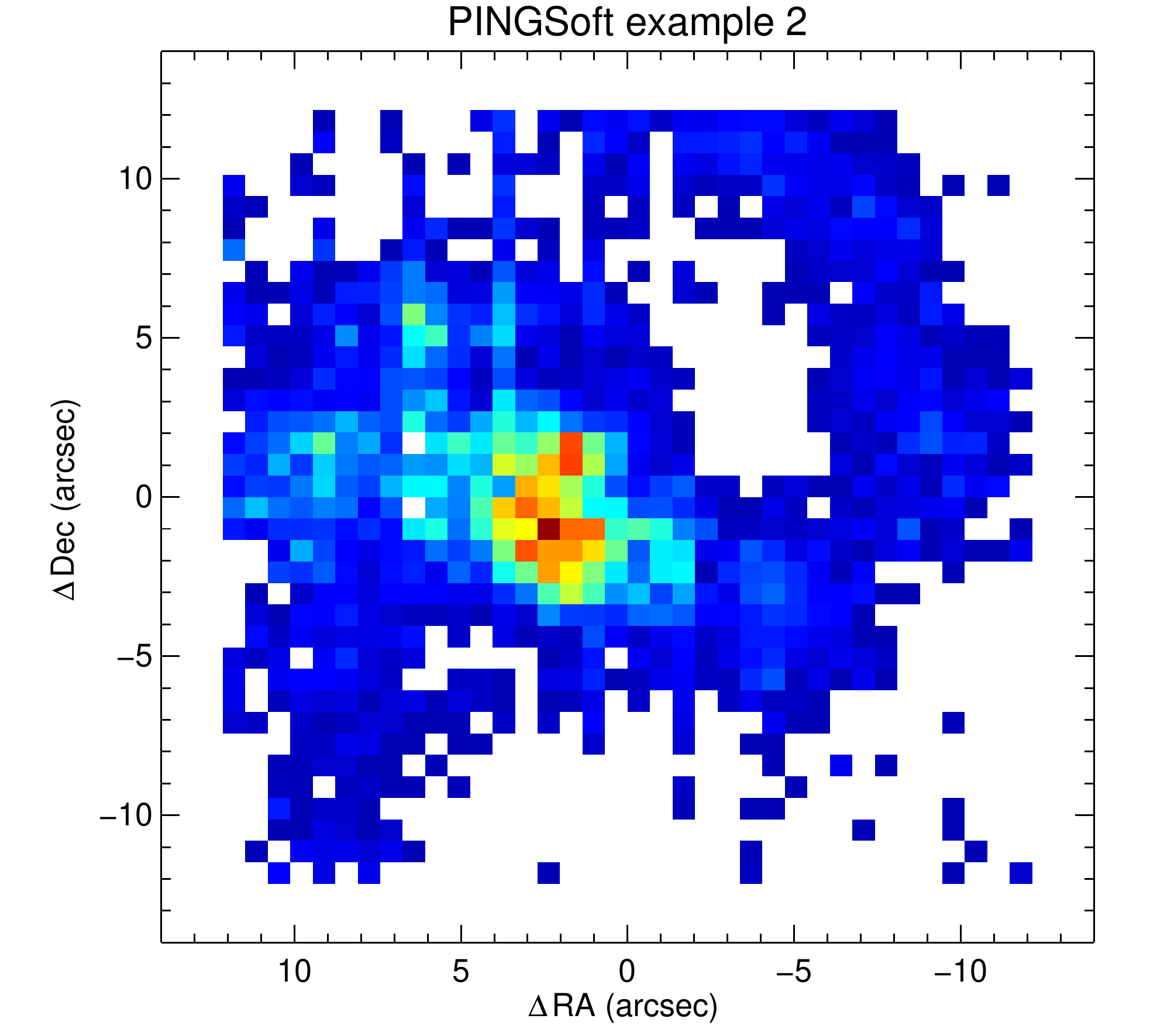}
    \caption[]
    {Postscript outputs examples of the {\tt view\_3D} routine
      \label{fig:examples}
    }
  \end{center}
\end{figure}

\noindent {\bf Examples:}
\vspace{0.3cm}

\noindent
To visualise the RSS file {\tt IRAS06295.VIMOS.fits} with position table named
{\tt IRAS06295.VIMOS.pt.txt} (both included in {\tt pingsoft\_examples/})
limiting the intensity and wavelength range on the spectral window,
drawing two vertical lines at lambda 6200 and 6700, using an inverse
hyperbolic sine function scaling:

{\footnotesize
\begin{verbatim}
view_3d, 'IRAS06295.VIMOS.fits',  min=-1, max=10, lmin=6200, lmax=6800, vline=[6200,6700], /asinh
\end{verbatim}
}
\vspace{0.3cm}

\noindent
To create a Postscript image of the {\tt ngc4625.dither.fits} RSS file, with
the spaxels drawn, the PINGSoft-special colour table, a 2\% clipping scaling,
using a narrow-band H$\alpha$ 20 \AA\ filter, and a special title (shown in
\autoref{fig:examples}):

{\footnotesize
\begin{verbatim}
view_3d, 'ngc4625.dither.fits', ct=44, filter=2, /clip, /draw, /ps, _extra={title:'PINGSoft example 1'}
\end{verbatim}
}
\vspace{0.3cm}

\noindent
To create a Postscript image of {\tt IRAS06295.VIMOS.fits}, with rainbow
colour table,  and using a B-Johnson (1965) filter and a special title (shown in
\autoref{fig:examples}):

{\footnotesize
\begin{verbatim}
view_3d, 'IRAS06295.VIMOS.fits', ct=33,  filter=3, /ps, _extra={title:'PINGSoft example 2'}
\end{verbatim}
}
\vspace{0.3cm}

\clearpage

{\large\noindent
The following routines are introduced in the {\sf PINGSoft} documentation.}\\

\subsection*{Spectra extraction}
\vspace{0.4cm}

\noindent
{\tt extract\_region}:
Extracts the spectra of regions selected by hand.
\vspace{0.2cm}

\noindent
{\tt extract\_aperture}:
Extracts the spectra within an elliptical or circular aperture.
\vspace{0.2cm}

\noindent
{\tt extract\_radial}:
Extracts radial average spectra within consecutive elliptical rings from
a reference point, based on either fixed bins or S/N floor.
\vspace{0.2cm}

\noindent
{\tt extract\_slit}:
Extracts the spectra within a rectangular aperture, resembling a long-slit
observation.
\vspace{0.2cm}

\noindent
{\tt extract\_cone}:
Extracts the spectra within a region defined by a hyperbolic cone.
\vspace{0.2cm}

\noindent
{\tt extract\_mask}:
Extracts the spectra based on a user's given mask or segmentation map.
\vspace{0.4cm}

\subsection*{Data products and analysis}
\vspace{0.4cm}

\noindent
{\tt extract\_filter}:
Generates a FITS image after convolving the 3D data with a narrow or broad-band filter.
\vspace{0.2cm}

\noindent
{\tt s2n\_ratio\_3D}:
Extracts spectra interactively based on a continuum and emission line S/N floor.
\vspace{0.2cm}

\noindent
{\tt s2n\_optimize}:
Extracts spectra interactively based on a S/N optimization.
\vspace{0.2cm}

\noindent
{\tt vfield\_3D}:
Calculates the intrinsic velocity field in 3D data using a wavelength cross-correlation.
\vspace{0.2cm}

\noindent
{\tt voronoi\_3D}:
Applies a Voronoi tessallation to the IFS data using the Voronoi binning
method by Cappellari \& Copin (2003), MNRAS, 342, 345
\vspace{0.4cm}

\subsection*{IFS manipulation}
\vspace{0.4cm}

\noindent
{\tt split\_califa}:
Extracts the FITS extensions for the CALIFA data.
\vspace{0.2cm}

\noindent
{\tt read\_rss}:
Reads a RSS FITS file and stores the data into an IDL vector.
\vspace{0.2cm}

\noindent
{\tt merge\_rss}:
Merges a list of RSS files into a single RSS file.
\vspace{0.2cm}

\noindent{\tt show\_hdr}:
Shows on screen the header of a FITS file, which can be written to an ASCII file.
\vspace{0.2cm}

\noindent{\tt write\_hdr}:
Adds or updates an entry in the header of a FITS file, using the
{\tt fxaddpar.pro} utility.
\vspace{0.2cm}

\noindent{\tt copy\_hdr}:
Copies the header of one FITS file to another, USE WITH CAUTION!
\vspace{0.2cm}

\noindent{\tt cube2rss}:
Converts a 3D FITS cube with dimensions $X$, $Y$, $\lambda$ to a RSS FITS file
plus an {\em ad hoc} position table in ASCII format.
\vspace{0.4cm}

\subsection*{Miscellaneous routines}

\noindent{\tt write\_wcs}:
Adds or updates the WCS (World Coordinate Systems) entries in a FITS header.
\vspace{0.2cm}

\noindent{\tt get\_new\_pt}:
Generates a new position table based on an index of selected spaxels.
\vspace{0.2cm}

\noindent{\tt shift\_ptable}:
Shifts the reference point or applies an offset to a given position table.
\vspace{0.2cm}

\noindent{\tt merge\_ptable}:
Concatenates a list of position table files into a single one for mosaicking
purposes.
\vspace{0.2cm}

\noindent{\tt offset2radec}:
Transforms small angle offsets in arcsec from a reference point to equatorial
coordinates.
\vspace{0.2cm}

\noindent{\tt radec2offset}:
Transforms equatorial coordinates to small angle offsets from a given
reference point.
\vspace{0.2cm}

\bibliographystyle{mn2e}
\bibliography{mnras}

\begin{thebibliography}{}

\bibitem[\protect\citeauthoryear{Cappellari \& Copin}{Cappellari \&
  Copin}{2003}]{Cappellari:2003p4046}
Cappellari M.,  Copin Y.,  2003, MNRAS, 342, 345

\bibitem[\protect\citeauthoryear{Rosales-Ortega}{Rosales-Ortega}{2011}]{Rosale%
sOrtega:2011p3845}
Rosales-Ortega F.~F.,  2011, NewA, 16, 220

\end{thebibliography}

\vfill

\begin{minipage}{0.85\textwidth}
\addcontentsline{toc}{section}{Acknowledgements}
\label{acknowledgements}

\noindent 
IMPORTANT: \\\vspace{-5pt}

\noindent
If you find this code useful for your research please acknowledge
the use of {\sf PINGSoft} in your publications:\\

\citet{RosalesOrtega:2011p3845} NewAstron 16, 220\\

\noindent 
Bugs, errors and inconsistencies (especially with non-tested
instruments) are expected. If you want to report a bug, or if you have any
comments or suggestions please contact the author at: \url{frosales@inaoep.mx}\\

\vspace{2cm}

\addcontentsline{toc}{section}{Copyright}
\label{license}

\noindent Copyright {\copyright} 2010, 2012  F. Fabián Rosales-Ortega\\

\noindent 
{\sf PINGSoft} is licensed under \hyperref[license]{GPLv3}.\\

\noindent
{\sf PINGSoft} is free software: you can redistribute it and/or modify
it under the terms of the GNU General Public License as published by
the Free Software Foundation, version 3.\\

\noindent 
{\sf PINGSoft} is distributed in the hope that it will be useful,
but WITHOUT ANY WARRANTY; without even the implied warranty of
MERCHANTABILITY or FITNESS FOR A PARTICULAR PURPOSE. See the
GNU General Public License for more details.\\

\noindent 
The GNU General Public License is found in: \url{http://www.gnu.org/licenses/gpl.html}

\end{minipage}

\end{document}